\documentclass[amsmath,amssymb,prb,aps,twocolumn,superscriptaddress,longbibliography,floatfix]{revtex4-2}
\usepackage{mathtools,bm,xcolor,graphicx,dcolumn}

\usepackage{hyperref}
\usepackage[capitalise]{cleveref}
\usepackage[version=3]{mhchem}
\usepackage{ulem}

\newcommand{\OIST}{Quantum Information Science and Technology Unit, Okinawa Institute of Science and Technology Graduate University, Onna-son, Okinawa 904-0495, Japan}

\newcommand{\UQ}{School of Mathematics and Physics, The University of Queensland, 4072, Australia}

\newcommand{\USYD}{School of Chemistry, University of Sydney, NSW 2006, Australia}

\begin{document}

\title{
Competing quantum effects in spin crossover chains: spin-orbit coupling, magnetic exchange, and elastic interactions
}

\author{F. Rist}
\affiliation{\UQ}
\author{H. L. Nourse}
\thanks{Present address: \USYD}
\affiliation{\OIST}
\author{B. J. Powell}
\affiliation{\UQ}

\date{\today}

\begin{abstract}
We derive and study a model of square planar, d$^8$ spin crossover materials that treats elastic, magnetic and spin-orbit interactions on an equal footing.  
For  1D chains density matrix renormalization group calculations show that the competition between these interactions leads to six different phases.
For weak spin-orbit coupling (SOC) and large antiferromagnetic interactions we find a symmetry protected topological (SPT)  Haldane phase. Moderate SOC introduces dynamical low-spin (LS) impurities that rapidly destroy the SPT state and drive the system into a topologically trivial high spin (THS) phase. This is equivalent to the Haldane--large-D phase transition driven by single ion anisotropy ($D$) in the spin-one Heisenberg model. 
For strong SOC  the $S^z=\pm1$ HS states are high-energy excitations. Thus, the system can be understood as a transverse field Ising model with the SOC playing the role of the transverse field. Consistent with this, we find a quantum phase transition between the THS phase and a quantum disordered (QD) phase.
However, if the magnetic coupling is non-zero or the HS and LS states of a single molecule are non-degenerate the $Z_2$ (Ising) symmetry is broken and  the phase transition becomes a crossover.
Thus, the QD phase and the THS phases are adiabatically connected, as, equivalently, are the large-D phase of the spin-one Heisenberg model and the quantum disordered phase of the transverse field Ising model.
We also find a ferroelastic LS phase, an antiferroelastic phase, with alternating HS and LS complexes, and a dimer phase, which results from the competition between antiferromagnetic and antiferroelastic interactions. 
\end{abstract}
\maketitle

\section{Introduction}
%Introduces SC and cooperativity
Spin crossover (SC) materials hold enormous promise for applications in spintronics and molecular switches \cite{molecularswitch}. A SC molecule is a metal-organic complex with two stable spin states. This occurs due to the competition between ligand field splitting, which favors a low-spin (LS) state, and Coulomb repulsion (Hund's rule coupling), which favors a high-spin (HS) \cite{gutlich}. Although SC is a property of the molecule, collective spin-state switching phenomena, such as multistep transitions \cite{jace_structure_property_2020,paez_multistep,Jref1,Jref2,Jref3,Jref4,Jref5,Jref6,Jref7,Jref8,Jref9,Jref10,Jref11,Jref12,Jref13,Jref14,Jref15,Jref16,Jref17,Jref18,Jref19,Jref20,Jref21,Jref22,Jref23,Jref24,Jref25,Jref26,Jref27,Jref28,Jref29,Jref30,Jref31,Jref32,Jref33,Jref34,Jref35,Jref36}, collective switching \cite{modelmaterial}, light-induced spin-state trapping \cite{chem2},  thermal hysteresis \cite{brooker_2015}, and complex relaxation of excited states \cite{relaxation_expt_,nadeem_jacs,HAUSER1991275, HAUSER1999471, 
	PhysRevB.82.214106, enachescu2004nonexponential, mishra2008temperature,
	chastanet2018high, 
	kershaw2015decoupled, money2007interplay, enachescu2003non, 
	balde2013light,  
	https://doi.org/10.1002/chem.201000433, https://doi.org/10.1002/ejic.201701127} have been widely observed in SC collections.  More exotic phases, including Devil's staircases \cite{collet_devil_staircase}, dimer phases \cite{timm}, spin-state semtics \cite{ising}, spin-state ices \cite{jace_ice}, and other Coulomb phases \cite{Jace_coulomb}, have been proposed.

%Introduce elastic cooperativity
The most widely discussed cause of these collective phenomena in SC materials is the elastic interactions between molecules, which couple the spin-states of molecules due to the large ($\sim10\%$) change in metal-ligand bond lengths that accompany spin-state switching \cite{jace_structure_properties, tutorial_elastic_multistep,tutorial_sco, sco_electro_elastic}. Elastic coupling can be either antiferroelastic or ferroelastic, i.e., can either favor neighbors with different spin states or the same spin states, respectively\cite{jace_ice,jace_structure_properties}. Such interactions can give rise to long-ranged  ordering of spin states \cite{jace_structure_property_2020,modelmaterial,Jref1,Jref2,Jref3,Jref4,Jref5,Jref6,Jref7,Jref8,Jref9,Jref10,Jref11,Jref12,Jref13,Jref14,Jref15,Jref16,Jref17,Jref18,Jref19,Jref20,Jref21,Jref22,Jref23,Jref24,Jref25,Jref26,Jref27,Jref28,Jref29,Jref30,Jref31,Jref32,Jref33,Jref34,Jref35,Jref36,paez_multistep}. Classical models, including Ising-like and elastic models \cite{wajn_pick_sco,jace_structure_properties,sco_electro_elastic,tutorial_elastic_multistep,timm2,timm3,tutorial_sco,sco_electro_elastic,tutorial_elastic_multistep,boukheddaden_multistep,tutorial_elastic_multistep,jace_ice,jace_structure_properties,nadeem2}, have been widely invoked to explained this plethora of experimentally observed SC phenomena. 

However, SC materials can have important quantum effects arising from both magnetic exchange interactions \cite{Jeschke} and spin-orbit coupling (SOC) \cite{Spinorbit_buhks,SOC_tranverse_field,nadeem_jacs,nadeem2}. For example, density functional theory calculations show that Fe(II) triazole SC chains have magnetic spin exchange strengths of the same order as elastic coupling \cite{Jeschke}, yet these interactions are often either neglected or only included classically  \cite{sco_electro_elastic,tutorial_elastic_multistep,timm2,timm3,tutorial_elastic_multistep,blume}.
SOC is vital for SC as it allows (otherwise spin-forbidden) transitions between spin-states  \cite{Spinorbit_buhks,nadeem_jacs}.

Timm and Schollw\"ock  \cite{timm} have previously  studied  the interplay between SC and antiferromagnetic  exchange in one-dimensional chains of d$^6$ complexes, which have spin-zero LS states and spin-two HS states. For strong enough antiferromagnetic interactions they found that the ground state is in the (topologically trivial) $S=2$ Haldane phase \cite{pollman1}.  SC  enriches the possible ground states by introducing spin-$0$ dynamical impurities which favor a long-ranged ordered dimer phase that is a compromise between magnetic and spin-state ordering. 

However, Timm and Schollw\"ock did not investigate the effects of SOC in SC chains \cite{timm}. This means that, in their model, the spin-state (i.e., the total spin) of a complex is a constant of motion. Once SOC is included, this is not the case. We will show below that this has important consequences for the quantum many-body physics of SC chains. 

SOC is well known to be important to spin crossover with its effects well studied \cite{Spinorbit_buhks,nadeem_jacs,SOC_tranverse_field,nadeem2}. However, most work to date has focused on the effects of SOC in single SC molecules rather than its impact on collective effects. D'Avino \textit{et al}.  \cite{SOC_tranverse_field} carried out a detailed study of a phenomenological model of SOC in SC individual molecules. They also considered collective effects within the mean field approximation, noting that the SOC induces an effective transverse field in the Ising-like models of SC. However, they concluded that SOC only has ``minor effects on the thermodynamic properties and transition curves'' of SC materials. We will show below that this is not always the case. Nadeem \textit{et al}.  \cite{nadeem_jacs,nadeem2} included SOC in a many-body theory of SC based on crystal field theory, but integrated out quantum effects and limited their discussion to a semi-classical theory.

Recently, Liu \textit{et al}. \cite{modelmaterial} observed an antiferroelastic phase has been observed  in chains of the nickel tetrahydrobenzene (NiTHB) on Au(111). This contrasts with most other  SC chains, such Fe(II) 1,2,4-triazole, which show ferroelastic order. Density functional calculations suggest that the strain due to binding to the substrate may be important in stabilising the antiferroelastic phase at low temperatures. Consistent with this, at low temperatures NiTHB is in the LS phase on Au(100), where the Ni-Ni distance is smaller than on Au(111), and the HS phase on Au(110), where the Ni-Ni distance is larger than on Au(111) \cite{Liu21}. Liu \textit{et al}. also demonstrated dynamical collective spin-state switching between the HLHLHL$\dots$ and LHLHLH$\dots$ states, suggesting that such phases could be useful for data storage. While we will not attempt to model NiTHB specifically, this motivates us to study square planar SC complexes, which are highly compatible with growth on a surface. The d$^8$ filling, relevant to NiTHB, is also particularly interesting as SOC allows transitions between the ($S=1$) HS and ($S=0$) LS states at first order (in the SOC strength). This contrasts with the common example of Fe(II), which is d$^6$, where the ($S=2$) HS and ($S=0$) LS states are only coupled at second order via an ($S=1$) intermediate spin state \cite{Spinorbit_buhks,nadeem_jacs}. This suggests that the effects of SOC may be more prominent d$^8$ complexes.

\begin{figure}
	\begin{centering}
		\includegraphics[width=\linewidth]{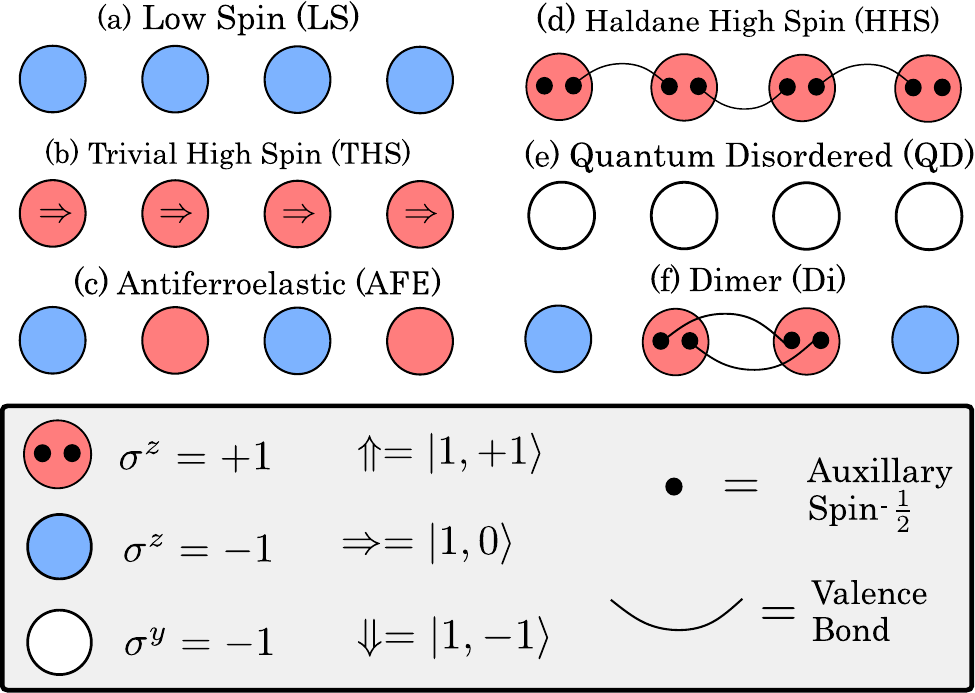}
		\caption{\label{fig:phases}
			Schematic of the groundstates found in our simple model of square planar, d$^8$, spin crossover complexes with magnetic exchange and spin-orbit coupling [\cref{eqn:Ham}]. The colour of a site represents the spin-state (pseudo-spin), with red corresponding to spin-one, high-spin (HS; $\sigma^z = 1$), blue to spin-zero, low-spin (LS; $\sigma^z = -1$), and white to a  superposition of HS and LS ($\sigma^y = -1$). Arrows represent the $S^z$-projection of the spin. In the dimer and Haldane phases each spin-one can be represented by two spin-$1/2$s, which form  spin-singlets (valence bonds) with the spin-$1/2$s on neighboring sites.
		}
	\end{centering}
\end{figure}

This Article is organised as follows. In \cref{sec:model} we derive  an effective low-energy model of SC in d$^8$ square planar spin crossover complexes.
This model treats  elastic interactions, SOC, and magnetic exchange on an equal footing. 
We show that SOC gives rise to a term that behaves as a single ion anisotropy for the physical spin and a transverse field for the pseudo-spins that label the spin-states of the molecules.  
In \cref{sec:methods} we describe our density matrix renormalisation group (DMRG) calculations for this model. 
In  \cref{sec:phase-diagrams} we report the results of these calculations. They reveal a complicated zero temperature phase diagram containing six phases.
We find that (anti)ferroelastic interactions favor (anti)ferroelastic  spin-state order   [specifically, the LS, \cref{fig:phases}a, trivial HS (THS) \cref{fig:phases}b; and  antiferroelastic (AFE), \cref{fig:phases}c, phases]. 
Exchange interactions favor a symmetry protected topological (SPT)  Haldane high spin (HHS) phase (\cref{fig:phases}d) \cite{haldane1983a,haldane1983b,wen2009spt,senthil2015spt,wen2017spt})
SOC favors a quantum disordered (QD) phase (\cref{fig:phases}e), similar to that found in the high-field limit of the transverse Ising model \cite{sachdev_2011}). 
There is strong competition between these interactions leading to a complex phase diagram, that also contains a  dimer phase (\cref{fig:phases}f), similar to that discovered by Timm and Schollw\"ock  \cite{timm}, which is compromise between the antiferroelastic and antiferromagnetic interactions.

\section{Model}\label{sec:model}

\subsection{Low-energy subspace of a SC complex}

\begin{figure}
	\centering
	\includegraphics[width=\linewidth]{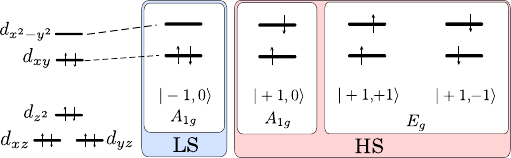}
	\caption{\label{fig:orbital}
		The orbital structure of a d$^8$ square planar spin crossover (SC) complex. The low-energy subspace consists of  the LS (a singlet, with both electrons in the $d_{xy}$ orbital) and the HS (a triplet, with one electron in the $d_{xy}$ orbital and another in the $d_{x^2-y^2}$ orbital). We denote these four spin states $|\sigma^z, S^z\rangle$, where $\sigma^z$ is the $z$ projection of a pseudeospin that labels the spin-state ($\sigma^z=-1$ for LS and $\sigma^z=1$ for HS) and $S^z$ is the $z$ projection of the physical spin. $A_{1g}$ and $E_g$ label the relevant irreducible representations of the $D_{4h}$ double group.
	}
\end{figure}

Square planar complexes typically have a large splitting between the $t_{2g}$ %($d_{xz}$, $d_{yz}$, $d_{z^2}$) 
and $e_g$ % ($d_{xy}$, $d_{x^2-y^2}$) 
orbitals of the metal ions. Thus, for d$^8$ complexes the $t_{2g}$ orbitals will be fully occupied and we need consider only the $e_g$ orbitals, \cref{fig:orbital}. The Aufbau principle predicts a LS singlet  state with both electrons in the $d_{xy}$ orbital, whereas Hund's first rule predicts a HS triplet with one electron in each of $d_{xy}$ and $d_{x^2-y^2}$ orbitals. Thus, these are the electronic configurations one expects to be energetically competitive in a square planar d$^8$ complex \cite{sugano_tanabe}. 

The simplest possible model therefore consists of a low-energy subspace in which each complex can take only these four possible states (i.e., the LS or one of the three HS states).
We denote the electronic state of the $i$th ion $|\sigma^z, S^z\rangle_i$, where the pseudo-spin $\sigma^z = -1$ ($+1$) indicates that the ion is LS (HS) and $S^z$ is the projection of the metal ion's spin onto the $z$-axis. That is, 
\begin{align}
	\label{eqn:basis}
	|-1,0\rangle_i & = \hat{c}^\dag_{i,xy,\uparrow} \hat{c}^\dag_{i,xy,\downarrow}|0\rangle,
	\nonumber \\
	|1,0\rangle_i & = \frac{1}{\sqrt{2}}(\hat{c}^\dag_{i,xy,\uparrow} \hat{c}^\dag_{i,x^2-y^2,\downarrow} + \hat{c}^\dag_{i,xy,\downarrow} \hat{c}^\dag_{i,x^2-y^2,\uparrow})|0\rangle ,
	\nonumber \\
	|1,1\rangle_i & = \hat{c}^\dag_{i,xy,\uparrow} \hat{c}^\dag_{i,x^2-y^2,\uparrow} |0\rangle,
	\nonumber \\
	|1,-1\rangle_i & = \hat{c}^\dag_{i,xy,\downarrow} \hat{c}^\dag_{i,x^2-y^2,\downarrow} |0\rangle,
\end{align}
where $\hat{c}^\dag_{i,\nu,s}$ creates an electron with spin $s$ in the $d_\nu$ orbital of the $i$th ion, and $|0\rangle$ is the vacuum of the low-energy subspace.

The creation operators for the $e_g$-orbitals can be rewritten as \cite{ballhausen}
\begin{equation}
	\label{eqn:basis_change}
	\begin{split}
		\hat{c}^\dag_{i,x^2-y^2,s} &= \frac{1}{\sqrt{2}}(\hat{a}_{i,2,s}^\dag + \hat{a}_{i,-2,s}^\dag ),\\
		\hat{c}^\dag_{i,xy,s} &= -\frac{i}{\sqrt{2}}(\hat{a}_{i,2,s}^\dag - \hat{a}_{i,-2,s}^\dag ),
	\end{split}
\end{equation}
where $\hat{a}^{(\dag)}_{i,{l_m},s}$ annihilates (creates) an electron on the $i$th ion with angular momentum $l_m$ and spin $s$.

\subsection{Wajnflasz--Pick--Ising model of spin crossover}

We model collective SC effects via the Ising-like model for pseudo-spins introduced by Wajnflasz and Pick \cite{wajn_pick_sco, tutorial_sco}. %This introduces a pesudospin variable  $\sigma^z_{j}=+1$ ($-1$) for HS (LS) states. 
The elastic interactions are described via a phenomenological parameter, $J_I$. We will describe the coupling as (anti)ferroelastic for ($J_I > 0$) $J_I < 0$.
In general the different spin-states of an isolated SC complex will have different  free-energies. This can be re-summed and included in the many-body Hamiltonian, whence the free-energy difference between HS and LS,  $\Delta G$, appears as an effective longitudinal field in the Ising-like model. For a chain of $N$  complexes the model is
\begin{equation}
	\hat{H}_\text{WPI}=J_I\sum_{j=1}^N {\sigma}^z_j{\sigma}^z_{j+1}+ \Delta G\sum_{j=1}^N{\sigma}^z_j.
\end{equation}

\subsection{Magnetic exchange}

Magnetic coupling between nearest neighbor metallic ions is facilitated by superexchange, mediated through the ligands: 
\begin{equation}
	\hat{H}_{\mathrm{X}}= J_H\sum_{j=1}^N
	\hat{\textbf{S}}_j\cdotp \hat{\textbf{S}}_{j+1},
\end{equation}
where $\hat{\mathbf{S}}_i=(\hat{{S}}_i^x,\hat{S}_i^y,\hat{S}_i^z)$ is the spin operator for the $i$th complex.
We only consider antiferromagnetic coupling ($J_H > 0$) in accordance with the Goodenough-Kanamori rules \cite{Kanamori_1959}. This expectation has been confirmed by first principles calculations for SC compounds \cite{Jeschke,copper_triazole}, which also suggest that magnetic interactions are non-negligible in these materials.

\subsection{Spin-orbit coupling \label{sec:soc}}

Spin-orbit coupling allows for inter-system crossing between electronic states, facilitating spin-state switching \cite{Spinorbit_buhks, nadeem_jacs}. This is given by
\begin{equation}
\label{eqn:soc_generic}
\hat{H}_{\mathrm{SOC}}=\xi \sum_{j=1}^N \hat{\mathbf{L}}_j \cdot \hat{\mathbf{S}}_j,
\end{equation}
where $\hat{\mathbf{L}}_i=(\hat{{L}}_i^x,\hat{{L}}_i^y,\hat{{L}}_i^z)$ is the orbital angular momentum operator on the $i$th complex and $\xi$ is the spin-orbit coupling strength.

When spin-orbit coupling is introduced, the symmetry of the metallic ion is described by the $D_{4h}$ double group. The LS state is a representation of $A_{1g}$, as is  the $S^z = 0$ HS state, while the $S^z = \pm 1$ HS states form an $E_g$ irreducible representation. Hence, only the LS state and the $S^z=0$ HS state are coupled. This results in a $\Delta S^z=0$ selection rule within the low-energy subspace.   %We only consider spin-orbit coupling up to first-order due to the relatively small coupling strength for transition metals, particularly for square-planar nickel(II) complexes \cite{SOC_nickel_anomaly}. 
 Thus, we find that
\begin{equation}
\begin{split}
\hat{H}_{\mathrm{SOC}}& = \xi \sum_{j=1}^N  \hat{L}_j^z \hat{S}_j^z,\\
&= \frac{\xi}{2} \sum_{j=1}^N \sum_{s\in\{\downarrow,\uparrow\}}\sum_{l_m=-2}^2(-1)^{\delta_{s,\downarrow}}\ l_m\hat{a}_{j,l_m,s}^\dag \hat{a}_{j,l_m,s},  \\
& = i \xi \sum_{j=1}^N  \bigg{(}\hat{c}^\dag_{j,xy,\uparrow}\hat{c}_{j,x^2-y^2,\uparrow} 
-\hat{c}^\dag_{j,xy,\downarrow}\hat{c}_{j,x^2-y^2,\downarrow}
\\
& \phantom{=} + \hat{c}^\dag_{j,x^2-y^2,\downarrow}\hat{c}_{j,xy,\downarrow}
-\hat{c}^\dag_{x^2-y^2,\uparrow}\hat{c}_{xy,\uparrow}\bigg{)},\\
& =-i \sqrt{2}\xi\sum_{j=1}^N\big(|1,0\rangle_j  {_j}\langle -1,0|
%\\ & \phantom{=} 
-|-1,0\rangle_{j} {_j}\langle 1,0|\big).
\end{split}
\label{eqn:soc_sco_basis}
\end{equation}

It is convenient to introduce pseudo-spin ladder operators, defined as
\begin{align}
%\hat{\sigma}_j^+ |\sigma^z,S^z\rangle_j = \sqrt{2-\sigma^z(\sigma^z+1)} |\sigma^z+2,S^z\rangle_j,
%\nonumber \\
%\hat{\sigma}_j^- |\sigma^z,S^z\rangle_j = \sqrt{2-\sigma^z(\sigma^z-1)}|\sigma^z-2,S^z\rangle_j.
\hat{\sigma}_j^+ |\sigma^z,S^z\rangle_j = \left(\frac{1-\sigma^z}2\right) |\sigma^z+2,S^z\rangle_j,
\nonumber \\
\hat{\sigma}_j^- |\sigma^z,S^z\rangle_j = \left(\frac{1+\sigma^z}2\right) |\sigma^z-2,S^z\rangle_j.
\end{align}
Whence the spin-orbit coupling can be written as
\begin{equation}
		\hat{H}_{\mathrm{SOC}}  =\lambda \sum_{j=1}^N \hat{\sigma}^y_j [ 1-(\hat{S}^z_j)^2] ,
	\label{eq:sco_effective_ham}
\end{equation}
where $\lambda = \sqrt{2}\xi$, $\hat{\sigma}^y_j = -i( \hat{\sigma}^+_j - \hat{\sigma}^-_j )$, and the factor $1-(\hat{S}^z)^2 = \delta_{S^z,0}$ enforces the $\Delta S^z=0$ selection rule. We only consider the case of $\lambda>0$ as the spin-orbit coupling constant will be positive for nickel in a square planar geometry \cite{SOC_nickel}.

Some insight into the physics engendered by the SOC can be gained by assuming simple orders in one of the spin or pseudo-spin subsystems.
If $\langle\hat{S}^z_j\rangle=0$ for each ion, the SOC acts like a  transverse field for the pseudo-spins, similarly to a previously proposed model for SC \cite{SOC_tranverse_field}.
Conversely, if $\langle\hat{\sigma}^y_j\rangle=\pm1$, the SOC acts like a single ion anisotropy for the physical spins.
\subsection{Full model}

Finally, we promote the classical pseudo-spin variable in the Wajnflasz--Pick--Ising model to a quantum operator, ${\sigma}^z_j\rightarrow\hat{\sigma}^z_j$. Summing the three terms described above yields our full model of a magnetically coupled chain of d$^8$  square planar SC complexes:
\begin{equation}
	\begin{split}
	\label{eqn:Ham}
	\hat{H}=&\hat{H}_{\mathrm{WPI}}+\hat{H}_{\mathrm{X}}+\hat{H}_{\mathrm{SOC}},\\
	=&J_I\sum_{j=1}^N\hat{\sigma}^z_j\hat{\sigma}^z_{j+1}+ \Delta G\sum_{j=1}^N\hat{\sigma}^z_j 
	\\ 
	&+ J_H\sum_{j=1}^N
	\hat{\textbf{S}}_j\cdotp \hat{\textbf{S}}_{j+1}+\lambda\sum_{j=1}^N\hat{\sigma}^y_j [1-(\hat{S}^z_j)^2].
	\end{split}
\end{equation}
An important difference between this model and many other  model Hamiltonians for SC systems is that the total spin is not a good quantum number for $\lambda\neq 0$.

%Below we describe how these interactions arise in SC materials, and in \cref{sec:soc} we derive the SOC term.

%%%%%%%%%%%%%%%%%%%%%%%%%%%%
%%%%%%%%%%%%%%%%%%%%%%%%%%%%

%%%%%%%%%%%%%%%%%%%%%%%%%%%%
%%%%%%%%%%%%%%%%%%%%%%%%%%%%

\section{Methods}\label{sec:methods}

\subsection{Density matrix renormalization group calculations}

We solve for the groundstates of \cref{eqn:Ham} using DMRG \cite{white1992dmrg,white1993dmrg,schollwock2005dmrg} in the matrix product state (MPS) formalism \cite{rommer1997mps,dukelsky1998mps,mcculloch2007mps,schuch2008mps,oseledets2011mps,schollwock2011mps} using the ITensor library \cite{itensor}. We include the LS and three HS states by using a local $4$-dimensional Hilbert space and construct the relevant local operators in the Hamiltonian of the full model [\cref{eqn:Ham}]. We keep a truncated basis size of the tensors of $m=700$ for the majority of parameters in the phase diagram, using a smaller or larger $m$ depending on how close the ground state is to a phase boundary. We scale all our results to the $m\rightarrow \infty$ limit using the variance in the internal energy. We study lattice sizes up to $N=200$ and scale all of our observables to the thermodynamic limit $N\rightarrow \infty$.

\subsection{Order parameters}\label{sec:phase_summary}

Several order parameters are required to differentiate the six phases sketched in \cref{fig:phases}. For ease of reference we detail them below, before moving on to our results.

\subsubsection{(Anti)ferroelastic phases} \label{sec:pseudo_spin_dof}

Long-range spin-state order can be identified with local order parameters. The ferroelastic order parameter is analogous to magnetization:
\begin{equation}
\label{eq:mz}
m^z=\frac{1}{N}\sum_{j=1}^N \langle \hat{\sigma}^z_j\rangle .
\end{equation}
$m^z > 0$ ($m^z < 0$) indicates a HS (LS) phase, \cref{fig:phases} a,b,d. 

The antiferroelastic order parameter is also analogous to staggered magnetization:
\begin{equation}
\label{eq:mza}
m^z_A=\frac{1}{N}\sum_{j=1}^N (-1)^j\langle \hat{\sigma}^z_j\rangle .
\end{equation}
$m^z_A \ne 0$ indicates long-ranged  antiferroelastic order with alternating HS and LS, \cref{fig:phases}c. 

It will also be helpful to consider a transverse ferroelastic order parameter which measures the ordering described in \cref{fig:phases}e,
\begin{equation}
\label{eq:my}
m^y=\frac{1}{N}\sum_{j=1}^N \langle \hat{\sigma}^y_j\rangle .
\end{equation}

\subsubsection{Dimer ordering} \label{sec:dimer}
\noindent
The dimer phase can be identified by the parameter,
\begin{equation}
\label{eq:dimer}
\begin{split}
m^D=&\frac{1}{2N}\sum_{\alpha=1}^{3}\sum_{j=1}^N (-1)^{\delta_{\alpha,j  \% 3}}\langle\hat{\sigma}^z_j\rangle\\&\times\bigg{( }1-\delta_{\alpha,(j+1)  \% 3}+(-1)^{\delta_{\alpha,(j+1) \% 3}}\langle\hat{\textbf{S}}_j\cdotp\hat{\textbf{S}}_{j+1}\rangle\bigg{)},
\end{split}
\end{equation}
where  
\begin{equation}
\delta_{\alpha,j  \% 3} = \frac{1}{3}\left\{4\cos\left[\frac{2\pi}{3}(2j+\alpha-1)\right]-1\right\}.
\end{equation}
Although this is somewhat mathematically cumbersome the physical interpretation is straightforward: $m^D$ identifies long-range ordered phases with a repeating pattern of one LS complex and then two HS complexes in a spin singlet, as sketched in \cref{fig:phases}f.

\subsubsection{Haldane phase} \label{sec:spindof}

The Haldane phase is well known from previous work on integer spin Heisenberg models \cite{haldane1983a,haldane1983b}. For odd-integer spins it has SPT order and fractionalized  edge states (spin-$1/2$ for spin-one models) \cite{AKLT,pollman1}. In contrast to the long-ranged  order, an SPT state order cannot be explained by Ginzburg-Landau spontaneous symmetry breaking and cannot be characterized by local order parameters \cite{wen2009spt,senthil2015spt,wen2017spt}.  However, the SPT order of the Haldane phase can be protected by any of the following symmetries: dihedral ($D_2=Z_2\times Z_2$), time-reversal, or reflection about a bond \cite{wen2009spt,pollman1}. Dihedral symmetry leads to the formation of hidden non-local magnetic order \cite{AKLT_string}. This hidden order can be identified with the non-local trivial, 
\begin{equation}
	\label{eq:trivial-string}
	O^i=\lim_{|j-k|\rightarrow\infty} \bigg{\langle} \hat{\mathbb{I}}_j\exp(i\pi\sum_{l=j+1}^{k-1}\hat{S}^z_l)\hat{\mathbb{I}}_k \bigg{\rangle} , 
\end{equation} 
and  spin,
\begin{equation}
	\label{eq:spin-string}
	O^z=-\lim_{|j-k|\rightarrow\infty} \bigg{\langle} \hat{S}^z_j\exp(i\pi\sum_{l=j+1}^{k-1}\hat{S}^z_l)\hat{S}^z_k \bigg{\rangle} ,
\end{equation} 
string order parameters \cite{string_order},  where $\hat{\mathbb{I}}_j$ is the identity operator on site $j$. If $O^z\neq0$ and $O^i=0$ then the state is in the Haldane phase, whereas if $O^z=0$ and $O^i\neq0$ there is no hidden string order and the phase is topologically trivial   \cite{pollmann2012detection}.

%We also rely on other indicators for the presence of the spin-one Haldane phase such as the existence of fractionalised edge spin-$\frac{1}{2}$s \cite{} and an even-fold entanglement spectrum \cite{}. The destruction of the latter of these two indicators requires the breaking of time reversal, bond inversion and $D_2$ symmetries and would result in the destruction of the Haldane phase. This makes this measurable an excellent indicator whether a quantum state is in the Haldane phase or not \cite{pollman1}.

%a transition to the large $D$ phase of the anisotropic spin-$1$ Heisenberg modelc). %In the limit that $J_H/|J_I| \rightarrow 0$, $\lambda / |J_I| \rightarrow \infty$, and $\Delta G / |J_I| \rightarrow -\infty$, 

\section{Results\label{sec:phase-diagrams}}

\subsection{Ferroelastic interactions (\texorpdfstring{$J_I<0$}{})}\label{sec:ferro}

\subsubsection{Degenerate HS and LS states ($\Delta G=0$)}

\begin{figure*}
\begin{centering}
\includegraphics[width=\linewidth]{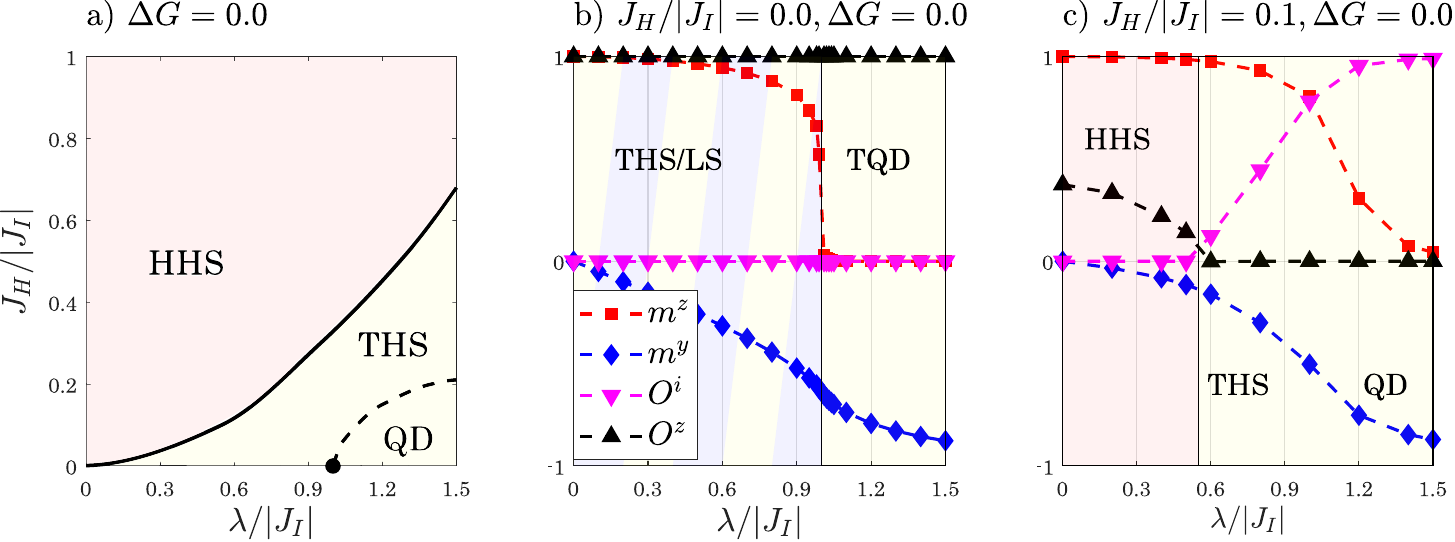}
\caption{\label{fig:FM}
(a) Phase diagram and (b),(c) order parameters for ferroelastic interactions ($J_I<0$) and degenerate spin states ($\Delta G=0$). Spin-orbit coupling ($\lambda$) destroys the singlets in the Haldane HS (HHS) phase, leading to a quantum phase transition to a trivial HS (THS) state. Further increasing $\lambda$ drives a crossover where the spin states become quantum disordered (QD). This phase is adiabatically connected to the large single ion anisotropy (large-$D$) phase found in spin-one chains.
$J_H$ is the antiferromagnetic spin exchange between adjacent sites and $J_I < 0$ is a ferroelastic coupling between sites. 
In panel (a) the circle denotes a quantum critical point, the phase boundary between the HHS and THS phases is denoted with a solid line, while the dashed line is a guide to the eye marking the crossover from THS to QD. In panels (b,c)  $m_z$ ($m_y$) is the longditudinal (transverse) ferroelastic order parameter defined in \cref{eq:mz} (\cref{eq:my}) and ($O^i$) $O^z$ is the (trivial) string order parameter given by (\cref{eq:trivial-string}) \cref{eq:spin-string}. 
In the absence of magnetic exchange ($J_H=0$; panel (b)) the THS and LS phases are degenerate ground states for weak SOC. There is a true quantum phase transition between the THS/LS phase and the QD phase at $\lambda=J_I$, which can be understood straightforwardly as corresponding to the quantum phase transition in the transverse field Ising model. Non-zero $\Delta G=0$ (\cref{sec:ferroDGne0}) or $J_H=0$ (panel c) lift the degeneracy between the HS and LS states on individual complexes and drive the quantum phase transition into a crossover.
}
\end{centering}
\end{figure*}

% What happens to the left of the critical point
For $\Delta G = 0$ and ferroelastic coupling we find three phases: the Haldane high spin (HHS), a topologically trivial trivial high spin (THS) and quantum disordered (QD) phases, \cref{fig:FM}. 

For weak SOC we find that $O^i =0$ and $O^z \neq 0$, indicating the Haldane phase. Indeed as $\lambda / J_H \rightarrow 0$ we find $O^z \approx 0.374$, the value found for the pure Heisenberg model \cite{White_DMRG_heisenberg}. As $\lambda / J_H$ increases, $O^z$ monotonically decreases to zero until there is a topological quantum phase transition to the THS phase, \cref{fig:FM}c. 

Throughout the HHS phase there is an extremely large fraction of HS complexes, $m^z \approx 1$, indicating that the magnetic correlations in the HHS phase are incompatible with dynamic LS impurities. When SOC is strong enough to induce even a very small fraction of LS complexes this rapidly suppresses and destroys the spin string order, and hence the SPT phase. Concomitantly, the large energy cost associated with breaking valence bonds between neighboring HS complexes in the HHS phase protects it against the appearance of LS complexes, stabilizing the phase.

% What happens to the right of the critical point
Once the HHS phase is destroyed, increasing $\lambda$ monotonically increases $O^i$ towards one (\cref{fig:FM}c). Simultaneously, $m^z\rightarrow0$ and $m^y\rightarrow-1$. This occurs because SOC favors a trivial quantum product state with $\hat{S}^z_i=0$ and $\hat{\sigma}_j^y=-1$ for all sites (see \cref{eq:sco_effective_ham}). Explicitly, for $\lambda  \rightarrow \infty$, $O^i \rightarrow 1$ and the ground state is
\begin{equation}
 |QD\rangle \equiv \bigotimes_{j=1}^N \frac1{\sqrt{2}} \left( |1,0\rangle_j -i |-1,0\rangle_j \right) \ .\label{eq:QD}
\end{equation}
In the large $\lambda/J_H$ limit the $S^z=\pm1$ HS states are high-energy excitations and become irrelevant. Hence, the model reduces to a transverse field Ising model \cite{sachdev_2011} in this limit. $|QD\rangle$ is precisely the quantum disordered phase expected for this limit. %Consistent with these expectations we find a quantum phase transition between the THS and QD phases at $\lambda=J_I$ for $J_H=0$, \cref{fig:FM}b.

Physically, the QD state corresponds to a quantum superposition of the LS and $S^z=0$ HS states at every site. Our model does not include any coupling to the `environment' such as molecular vibrations or lattice phonons, or any external modes. It would be interesting, but beyond the scope of this work, to determine whether such couplings will decohere the superposition, and if so how quickly. 

Another interesting question about the QD phase, assuming it remains coherent long enough to be measured, is what signatures it would display. 
An important probe of the spin-state is metal-ligand bond length, which can be measured via x-ray crystallography. Let us define the metal-ligand bond length operator for the $j$th complex, $\hat{r}_{j}$, which takes the values $_j\langle 1,0| \hat{r}_{j}|1,0\rangle_j \equiv r_H$ and $_j\langle -1,0| \hat{r}_{j}|-1,0\rangle_j \equiv r_L$ when the complex is HS and LS, respectively, and has quantum coherent off-diagonal contributions $_j\langle 1,0| \hat{r}_{j} |-1,0\rangle_j\equiv r_{\mathrm{coh}}\equiv r_{\mathrm{coh}}^{\prime}+ir_{\mathrm{coh}}^{\prime\prime}$. It is important to note that x-ray crystallography measures the average bond length across the sample. A simple calculation yields
%\begin{widetext}
\begin{eqnarray}
 \frac1N\sum_{j=1}^N\left\langle QD\left|\hat{r}_{j}\right|QD\right\rangle 
    &=
   % \sum_{j=1}^N \frac{1}{2N}  \left( _j\langle 1,1,0| +i _j\langle-1,0,0| \right) \hat{r}_{j}
   % \left( |1,1,0\rangle_j -i |-1,0,0\rangle_j \right) \\
  %  &=& \sum_{j=1}^N \frac{1}{2N}  \left( r_H + r_L 
 %   -i _j\langle 1,1,0| \hat{r}_{j} |-1,0,0\rangle_j  +i _j\langle-1,0,0| \hat{r}_{j} |1,1,0\rangle_j   \right) \\
  %  &=& \sum_{j=1}^N \frac{1}{2N}  \left( r_H + r_L +
  %  2 r_{coh}^{\prime\prime}   \right)\\
    & \frac{r_H + r_L}{2}   + r_{\mathrm{coh}}^{\prime\prime}.
\end{eqnarray}
%\end{widetext}
As the metal-ligand bond length changes significantly between the LS and HS states of SC complexes we expect $r_{\mathrm{coh}}^{\prime\prime}$ will typically be small. Thus, x-ray crystallography will not distinguish the quantum disordered phase from classical disorder.

A similar analysis will hold for any other measurement that can be written as the sum of local terms. A second measurement common in SC is the  magnetic susceptibility, $\chi$, multiplied by temperature, $T$, given by $\chi^\alpha T=\sum_{j=1}^N \langle S^\alpha_{N/2}S^\alpha_j\rangle$  \cite{white_thermal_ancilla}. As there are no inter-complex magnetic correlations in a product state such as the QD state, $\chi T$ is a sum of local terms plus a coherent part. 
%the average for the all HS-chain and the all-LS chain. %The pure QD state is indistinguishable from LS phase from the perspective of magnetic susceptibility measurements, as both read zero. 
However, spin operators (such as $S^z_j$) do not connect  HS and LS states, thus the quantum coherent terms vanish. The expectation of any spin-operator for $|1,0\rangle_j$ and $|0,0\rangle_j$ will also be zero, resulting in a vanishing magnetic susceptibility for the ideal QD state [\cref{eq:QD}]. More generally, in the QD phase one should expect some  small population of the $|1,\pm1\rangle_j$ states. These will result in a concomitantly small, but non-zero, $\chi T$. In contrast, a classically disordered state of LS and all possible HS states will follow the Curie-Weiss law for a paramagnet, $\chi T \sim (1-\gamma_{HS}) S_{LS}(S_{LS}+1) + \gamma_{HS} S_{HS}(S_{HS}+1) =2\gamma_{HS}$, where $\gamma_{HS}$ is the fraction of HS complexes, $S_{LS}$ ($S_{HS}$) is the spin of the LS (HS) state, and we specialize to square planar $d^8$ complexes in the final equality.

Thus, the combination of crystallography and magnetic susceptibility measurements is sufficient to identify the QD phase.

At intermediate $\lambda / J_H$ the many-body physics is more subtle. To understand this regime it is helpful to first consider the model with $J_H=\Delta G=0$, \cref{fig:FM}b. Clearly, with no magnetic interactions between the physical spins the magnetic state is always trivial. We find that $m^y<0$ and $\langle (\hat S^z_i)^2\rangle=0$ for $\lambda>0$. Thus, we can understand the physics purely from the transverse field Ising model. This model is well known to have a quantum critical point at $\lambda=|J_I|$, from a ferroelastic spin-state ordered phase (for $\lambda<|J_I|$) to a quantum disordered phase  (for $\lambda>|J_I|$) \cite{sachdev_2011}. Unsurprisingly, our numerical results are in good agreement with this, \cref{fig:FM}b.

As well as the HHS phase at small $\lambda/J_H$, we also find a large region of the phase diagram where $O^z=0$ and $O^i>0$ but $m^z$ is large and positive while $m^y$ is small and negative, \cref{fig:FM}b. This indicates a topologically trivial HS ferroelastic (THS) phase.

If  $\langle \sigma^y_j\rangle$ is  small and negative, the SOC (\cref{eq:sco_effective_ham}) acts as a single ion anisotropy for the physical spins, often denoted $D(S^z)^2$. As increasing $\lambda$ increases $|m^y|$ this generates a super-linear increase in the effective single ion anisotropy.
Thus, the HHS to THS  phase transition in our model is analogous to the phase transition between the Haldane phase and the large-$D$ phase in the spin-$1$ Heisenberg model with single ion anisotropy \cite{wen2009spt}. 

Single ion anisotropy in a pure spin model arises from SOC, but appears only at second order in $\lambda$. In the SC chain we see a related effect, but now occurring at first order in $\lambda$. The large-$D$ phase has been widely studied theoretically and the Haldane--to--large-$D$ transition is a canonical example of a topological phase transition in the theoretical literature. There are materials known to realize the Haldane and large-$D$ phases \cite{Maximova2021a}, but none offer fine-tuning in order to explore the nature of the phase transition.
%This is presumably, at least in part, because SOC is usually weak and so $D$ is typically small compared to $J_H$ in magnetic materials.
Therefore, SC materials, where the equivalent transition is driven by a term at first order in $\lambda$, may be an interesting playground to explore this physics experimentally. 

The most obvious way to increase SOC is to move to heavier metals. However, this may be problematic here as SC is most commonly observed in complexes of third row transition metals. Nevertheless, this does not mean that SOC cannot be modified \cite{NadeemJACS}. In transition metal complexes  the effective spin-orbit coupling is strongly modified by the relativistic nephelauxetic effect \cite{Neese}, which is caused by a combination of: the covalent mixing of the metal and ligand wave functions; the expansion of the $d$-orbital of the metal in the process of forming a covalent bond with the ligand; and the SOC of the ligand. Thus, ligands with high covalency and large SOC will lead to larger $\lambda$.

In our extensive parameter search, we have found that the THS phase always appears as an intermediate phase between the HHS and QD phases for $\Delta G = 0$. SOC suppresses both ferroelastic and spin string order parameters. However, the spin string order is destroyed before the ferroelastic ordering is significantly weakened,  Fig. \ref{fig:FM}b,c. Consequently, to transition from the HHS phase (with $m^z\approx 1$) to the QD phase (with $m^z\rightarrow 0$) the system must pass through the THS phase ($1>m^z\gg0$).

There is clear numerical evidence (from the string order parameters, \cref{fig:FM}c that the change from HHS to THS is a continuous (quantum) phase transition. This is consistent with the general expectation on moving from an SPT phase to a topologically trivial phase \cite{wen2009spt}.
For $J_H=0$ there is a quantum phase transition at $|J_I| \simeq |\lambda|$.
However, for $J_H>0$ we see no evidence of a phase transition between the THS and QD phases numerically, \cref{fig:FM}c, and there is no theoretical requirement for a phase transition between them. Therefore, we conclude that the change between these regimes is a crossover. This can be understood because the magnetic interactions lower the energies of neighboring HS complexes, but do not affect LS complexes. Thus, exchange interactions  break the $Z_2$ symmetry of the pseudo-spins, 
driving the quantum phase transition into a crossover. (Formally, the $Z_2$ symmetry emerges only once the $S^z=\pm1$ HS states are projected out of the low-energy model; whence the unitary $\hat{U}\equiv\prod_{j+1}^N\hat{\sigma}_j^y$ maps $\hat{\sigma}_j^z \rightarrow -\hat{\sigma}_j^z$, leaving the Hamiltonian invariant.)  
This is consistent with the transverse Ising model where a longitudinal field turns the quantum phase transition into a crossover \cite{ising_renorm,Turner_interaction_distance}.

\subsubsection{Non-degenerate HS and LS states ($\Delta G\ne0$)}\label{sec:ferroDGne0}

\begin{figure*}
\begin{centering}
\includegraphics[width=\linewidth]{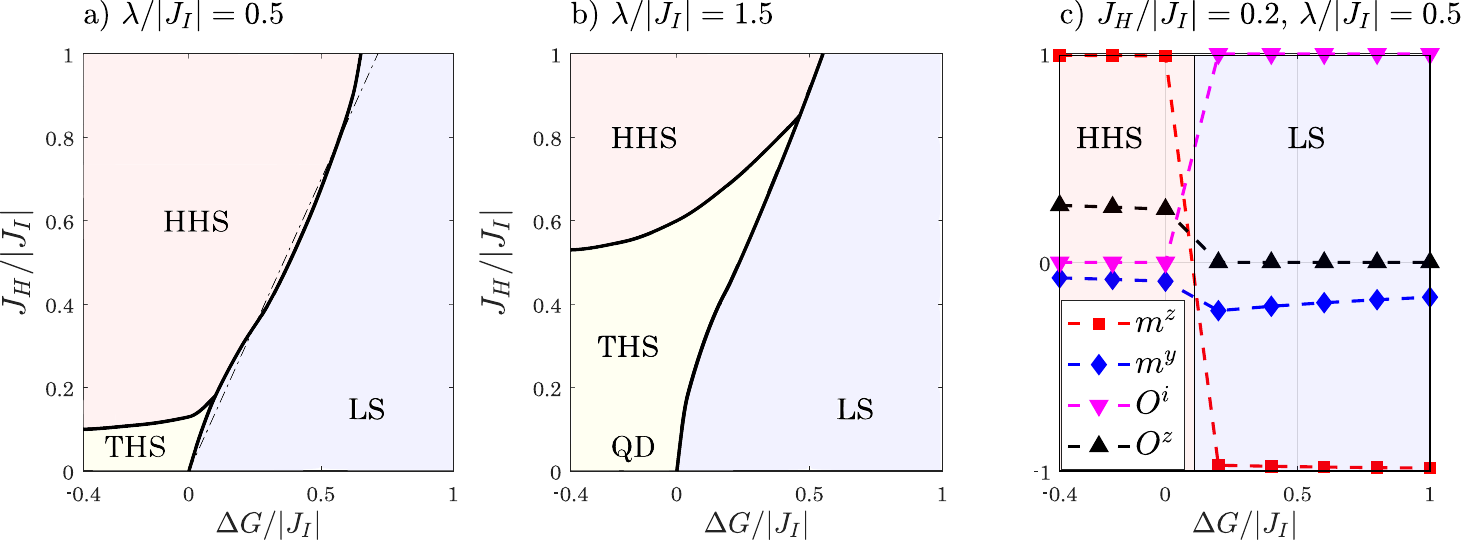}
\caption{\label{fig:FMG}
(a),(b) Phase diagrams and (b) order parameters for ferroelastic interactions with a non-zero Gibbs free energy difference $\Delta G=G_{HS}-G_{LS}$ between individual HS and LS molecules. For both (a) weak ($\lambda = 0.5J_I$) and (b) strong  ($\lambda = 1.5J_I$) spin-orbit coupling, increasing $\Delta G$  drives a phase transition to a  long-ranged LS  phase.
The dash-dotted line in (a) indicates $\Delta G = E_H$, the simple estimate for the HS-LS phase boundary neglecting $\lambda$ and $J_I$,  which is in remarkable agreement with the numerical data. 
The discontinuous change in order parameters as $\Delta G$ is varied (c) implies that there is no THS phase between the HHS and LS phases in contrast to the phase diagram for $\Delta G=0$ (\cref{fig:FM}).
}
\end{centering}
\end{figure*}

We do not expect any material to be fine tuned to $\Delta G=0$ at $T=0$. 
For  $\Delta G<0$ (individual HS complexes are lower energy than LS complexes) we find only the three phases present for $\Delta G=0$. But for sufficiently large $\Delta G>0$ a ferroelastic LS phase can be induced, \cref{fig:FMG}. For $\Delta G>0$  there is a competition between the single complex physics ($\Delta G$), which favors the LS phase, and antiferromagnetic spin exchange ($J_H$), which favors spin-singlets that can only occur between HS complexes. Thus, non-zero $J_H$ increases the critical $\Delta G$ for the HS-LS phase transition. For $J_H, \Delta G \gg |J_I|, \lambda$ the HS phase is the HHS phase; thus, the HS-LS phase transition moves to $\Delta G\simeq -E_H$, where $E_H \simeq-1.401J_H+\Delta G$ is the energy of the HHS phase. This result is in excellent agreement with our DMRG calculations, even when $\lambda$ is relatively large, \cref{fig:FMG}a.

For small $\lambda/J_I$  the  THS phase is pushed to smaller $J_H/J_I$, \cref{fig:FMG}a. This emphasises that the introduction of LS impurities, via  $\lambda$ or $\Delta G$, is essential to suppress the HHS phase.

A longitudinal field, $\Delta G\neq0$, explicitly breaks the $Z_2$ pseudo-spin symmetry. Consistent with this, our numerics suggests a crossover between THS and QD phases whenever $\Delta G\ne0$. Again, this follows our expectations from the transverse field Ising model with a longitudinal field \cite{ising_renorm,Turner_interaction_distance}.

For large  $\lambda/|J_I|$, \cref{fig:FMG}b, we also observe a larger region of THS phase as the HHS is suppressed by the introduction of LS impurities by the SOC.

\subsection{Antiferroelastic interactions (\texorpdfstring{$J_I>0$}{})} \label{sec:antiferro}

\subsubsection{Degenerate HS and LS states ($\Delta G=0$)}

\begin{figure}
\begin{centering}
\includegraphics[width=\linewidth]{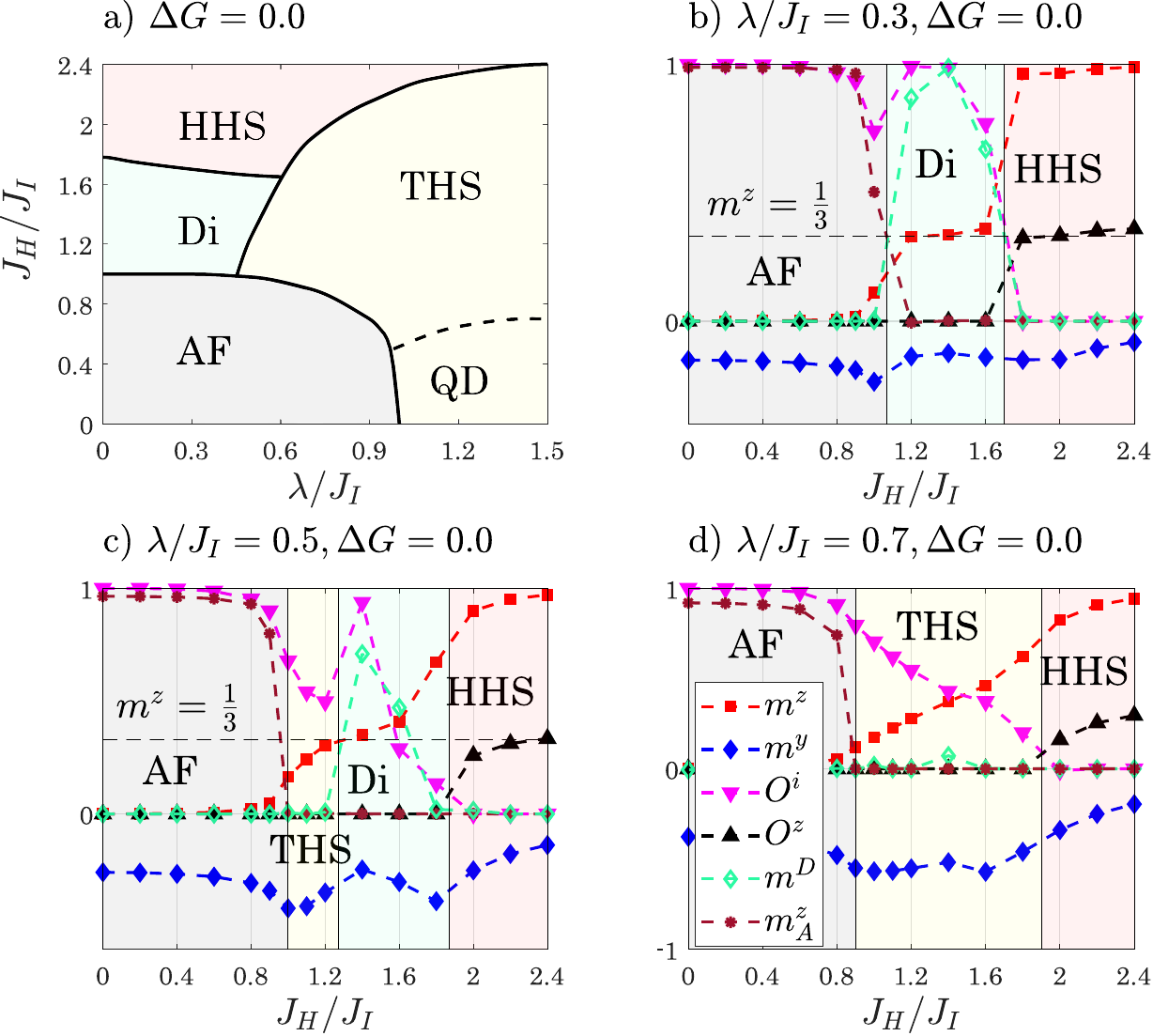}
\caption{\label{fig:AFEM}
(a) Phase diagram and (b-d) order parameters for antiferroelastic interactions ($J_I > 0$) and degenerate HS and LS complexes ($\Delta G = 0$).
Spin-exchange, which favors spin-singlet formation between HS complexes, competes with the antiferroelastic coupling and SOC; thus the HHS phase is limited to large $J_H$. For small $\lambda$ and intermediate $J_H$ there is a compromise and a long-ranged HS-HS-LS dimer (Di; \cref{fig:phases}f) order forms with adjacent HS complexes forming spin singlets, indicated by $m^D \neq 0$ (\cref{eq:dimer}) and $m^z \simeq 1/3$. For sufficiently large $\lambda$, the Di and AFE phases are suppressed due to disorder introduced by  dynamical LS impurities. 
} 
\end{centering}
\end{figure}

For antiferroelastic coupling we  find antiferroelastic (AFE; \cref{fig:phases}c) and dimer (Di; \cref{fig:phases}f) \cite{timm} phases, \cref{fig:AFEM} along with the  HHS, THS, and QD phases. For strong elastic interactions (small $J_H / J_I$ and $\lambda / J_I$) we find the AFE phase. Increasing $\lambda$ drives a  phase transition to the QD phase, which is adiabatically connected to the THS phase via a crossover.
As  $J_H / J_I$ is increased the stability of the AFE phase decreases and the phase boundary to the THS or QD phase is pushed to smaller $\lambda / J_I$ until the AFE phase disappears at $J_H / J_I \approx 1$.

The antiferromagnetic exchange ($J_H$) cooperates with ferroelastic interactions ($J_I < 0$) because if there are neighbouring HS states the system can lower its energy by forming spin singlets and developing long-ranged ferroelastic order. In contrast,  antiferroelastic interactions ($J_I > 0$) favor alternating HS and LS sites which destroys neighboring HS spin-singlets.
The Di phase is a compromise between these two  competing interactions. It is found for weak SOC (small $\lambda/J_I$) and  $J_H \simeq J_I$.

The Di phase  balances antiferromagnetic spin exchange and antiferroelastic coupling via long-ranged HS-HS-LS order, with the two adjacent HS forming a spin-singlet,  \cref{fig:phases}f. Thus, the Di phase is characterized by $m^z \approx 1/3$ and $|m_D| > 0$ (\cref{eq:dimer}; \cref{fig:AFEM}).  This should be evident from crystallography or M\"ossbauer spectroscopy. However, $\chi T$ should be small because of both the singlet formation as well as LSs.  Thus, combining susceptibility measurements with either crystallography or M\"ossbauer spectroscopy should provide a distinctive experimental signature of the Di phase.

Alternatively, experiments which measure spin spectral properties of the different phases such as neutron scattering and inelastic electron tunneling spectroscopy (IETS), which is performed with a scanning tunneling microscope, should also be able to identify different SC phases. On a metallic surface,  SC molecules in the HS state may show a Kondo resonance, whereas   LS molecules will not  \cite{modelmaterial}. This allowed the identification of  the AFE phase in NiTHB \cite{modelmaterial}. Similarly, the Haldane phase has previously been identified using IETS through zero-bias Kondo resonances produced by the fractionalised edge states \cite{Heinrich_IETS,Fransson_2010, theory_STM, Localprobes_heisenberg, Mishra_nanographene}. However, in the dimer phase the HS molecules form valence bonds and so the Kondo resonance zero bias peak should be absent, due to the monogamy of entanglement.

The discontinuous change in the antiferroelastic order parameter (\cref{eq:mza}) suggest that the AFE and Di phases are separated by a first order  phase transition, \cref{fig:AFEM}b.
For weak SOC and large enough $J_H/J_I$, spin-singlet formation wins and there is a  phase transition to the HHS phase. For $J_H\approx J_I$, increasing SOC drives a phase transition from the Di phase to the THS phase, Fig. \ref{fig:AFEM}a,c.

%what phases do we find when we consider AFE ising coupling?

%Phase transitions for the AFE as we increase SOC and superexchange

%\subsection{Antiferroelastic interactions ($J_I>0$) with a free energy difference (\texorpdfstring{$\Delta G \neq 0$}{})}\label{sec:antiferro_g}

\subsubsection{Non-degenerate HS and LS states ($\Delta G\ne0$)}

\begin{figure}
\begin{centering}
\includegraphics[width=\linewidth]{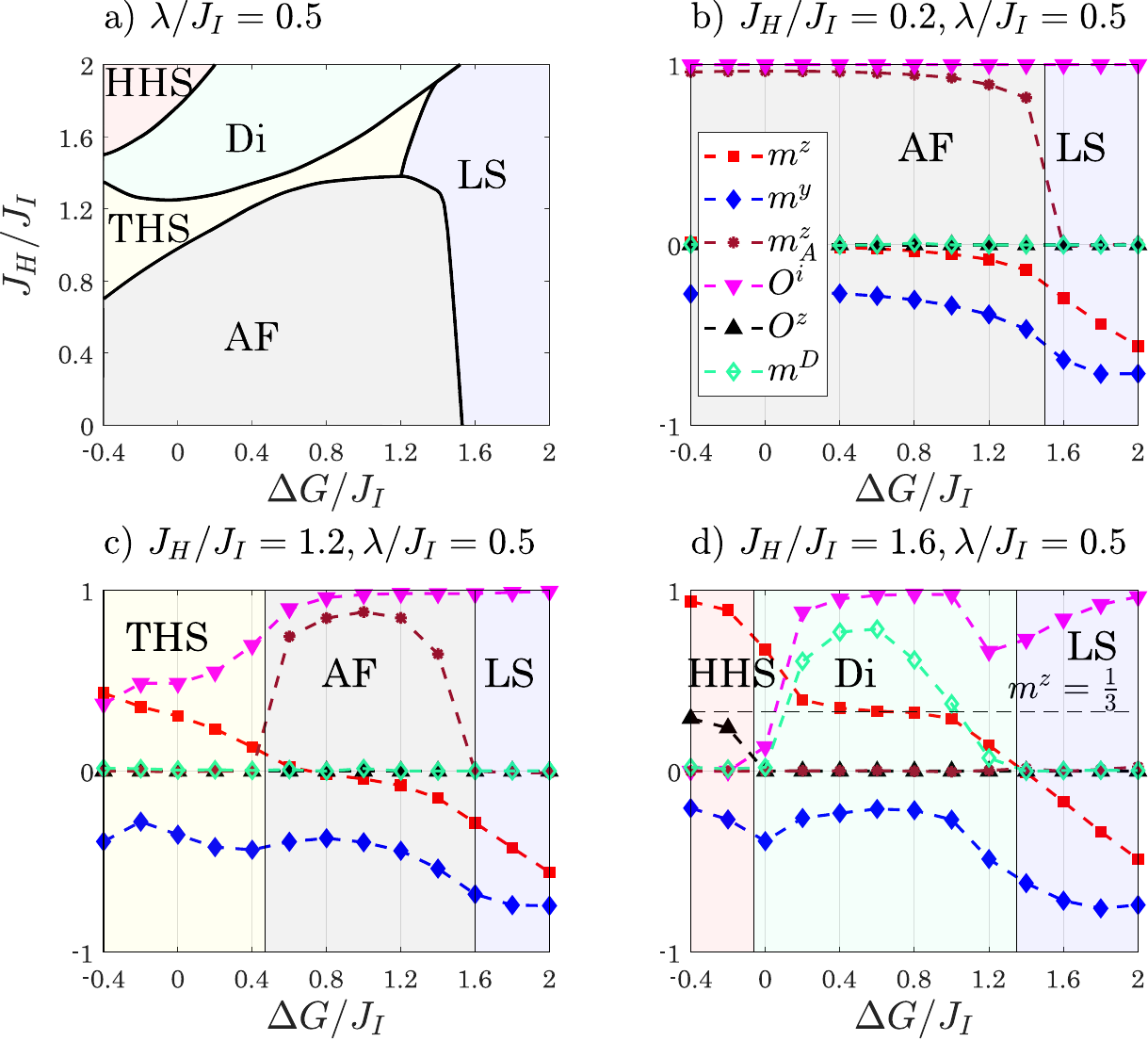}
\caption{\label{fig:AFEMG}
(a) Phase diagrams and (b-d) order parameters for antiferroelastic interactions ($J_I > 0$) and non-degenerate HS and LS complexes ($\Delta G \ne 0$) for weak SOC ($\lambda=0.5J_I$).
$\Delta G > 0$ stabilizes the AFE and Di phases as it favors  LS complexes. For large enough $\Delta G$ there is a phase transition to the long-ranged ordered LS  phase.
Antiferroelastic interactions suppresses both ferroelastic and dimer phases unless $\Delta G$ is  large.
}
\end{centering}
\end{figure}

For antiferroelastic interactions large enough $\Delta G > 0$ will drive LS long-ranged ferroelastic ordering ($m^z < 0$), similar to that found for ferroelastic coupling. 
However, the phase diagram in the antiferroelastic case, \cref{fig:AFEMG}, is significantly more complex than the phase diagram for ferroelastic interaction, \cref{fig:FMG}.

Notably, for weak SOC, the AFE phase is stable over a wide range of $\Delta G$ and weak to intermediate $J_H/J_I$. This is the phase observed in NiTHB \cite{modelmaterial}. 
For larger $J_H/J_I$, positive $\Delta G$ stabilizes the Di phase; although the HHS phase eventually wins out for very strong antiferromagnetic interactions.

There are now phase transitions between the AFE and LS phases, THS and LS phases, and Di and LS phases. However, unlike the ferroelastic case (\cref{fig:FMG}), we do not find a direct transition between the HHS and LS phases in our numerics, \cref{fig:AFEMG}. This indicates that the HHS phase is less resilient to LS impurities when there is antiferroelastic coupling. This is reasonable as we saw  that, even for ferroelastic interactions, the HHS phase is rapidly destroyed by the introduction of LS impurities, which  antiferroelastic coupling will tend to favor. Once even a small fraction of complexes become LS the Di phase is rapidly favoured as it is able to lower its energy from both antiferroelastic interactions (as two out of every three nearest neighbour are in opposite spin sites) and from antiferromagnetic exchange interactions, \cref{fig:phases}f.

%What happens to THS & TAFE when we introduce h GOOD

\begin{figure}
\begin{centering}
\includegraphics[width=\linewidth]{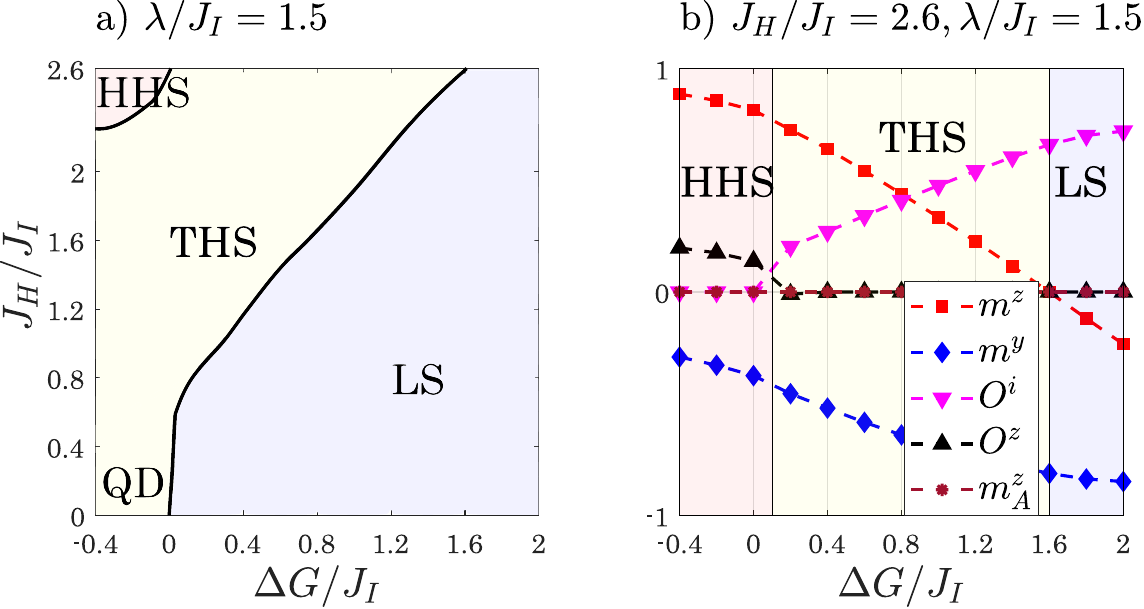}
\caption{\label{fig:AFMG15}
(a) Phase diagrams and (b) order parameters for 
antiferroelastic interactions with strong SOC, $\lambda/J_I=1.5$. Strong SOC suppresses both antiferroelastic and dimer phases and stabilises the QD phase as it promotes quantum disorder in the spin states. Nevertheless, there remains a competition between $\Delta G$ which favors long-range LS order and $J_H$ which favors long-range HS order. }
\end{centering}
\end{figure}

However, for strong SOC ($\lambda>J_I$) both the AFE and Di phases are suppressed, \cref{fig:AFMG15}, and the phase diagram is now broadly similar to that for ferroelastic interaction, \cref{fig:FMG}b. Interestingly $\Delta G>0$ suppresses the QD phase much more strongly than $\Delta G<0$. %as the QD phase is adiabatically connected to the THS phase, but not the LS phase.

\section{Conclusion}\label{sec:conclusion}

Elastic, magnetic and spin-orbit interactions have all been previously shown to significantly impact the collective behavior of spin crossover materials.  
We have derived and studied a model of spin crossover materials that treats all of these interactions on an equal footing.  
We solved the model for a 1D chain of square planar, d$^8$ complexes using finite DMRG at $T=0$. This showed a rich phase diagram containing six different phases (\cref{fig:phases}) due to the competition between these three different interactions.

SOC is key because it couples the magnetic and spin-state (pseudo-spin) degrees of freedom. For a given (fixed) magnetic state the SOC acts like a transverse field in the Wajnflasz--Pick--Ising model of pseudo-spins. Whereas for a given (fixed) configuration of spin states SOC acts as a single ion anisotropy in the spin-one Heisenberg chain. In  the round, SOC couples these two effects (\cref{eq:sco_effective_ham}). 
The inclusion of SOC  allow for dynamic changes of the spin-state of individual complexes, this leads to dynamic impurities and has a profound effect on the collective behaviors.

Because the HS state of square planar d$^8$ complexes is $S=1$ the HS phase comes in two forms: an SPT Haldane (HHS) phase and a topologically trivial HS (THS) phase. The Haldane phase is only stable for weak SOC and large antiferromagnetic interactions. This is because the dynamical LS impurities introduced for any sizable SOC rapidly destroy the SPT state and drive the system into the THS phase. For $\lambda>0$, which we expect physically, SOC favors $S^z=0$ states, thus the HHS-THS phase transition is equivalent to the Haldane--large-D phase transition driven by single ion anisotropy ($D$) in the spin-one Heisenberg model. 

Sizable SOC ($\lambda\gtrsim |J_I|$), degenerate HS and LS single molecule states ($\Delta G=0$), and vanishing magnetic interaction ($J_H=0$) implies that the $S^z=\pm1$ HS states are high-energy excitations and, therefore, do not play a significant role in the low-energy physics. 
The $S^z=\pm1$ HS states can therefore be projected out of the low-energy subspace, the resulting model has a $Z_2$ symmetry corresponding to  mapping $\hat\sigma^z_i\rightarrow-\hat\sigma^z_i$, generated by the unitary $\hat{U}\equiv\prod_{j+1}^N\hat{\sigma}_j^y$.
Thus, the system can be understood as a transverse field Ising model with the SOC, $\lambda$, playing the role of the transverse field. Consistent with this we find a quantum critical point between the THS and quantum disordered (QD) phases at $\lambda=J_I$ and $\Delta G=J_H=0$ numerically (without projecting the $S^z=\pm1$ HS states out of the model). These phases correspond to the ferromagnetic and quantum disordered phases of the transverse field Ising model respectively.
However, if either $\Delta G$ or $J_H$ are non-zero this lifts the $Z_2$ symmetry and changes the phase transition into a crossover.

The results above imply that there is adiabatic continuity between the QD phase and the THS phases in our model. More surprisingly, this implies that there is adiabatic continuity between the large-D phase of the spin-one Heisenberg model and the quantum disordered phase of the transverse field Ising model.

For $\Delta G>0$ we find a ferroelastic LS phase. Naturally, this is more stable for ferroelastic interactions than antiferroelastic interactions, as the former compete with the single ion physics, whereas the latter cooperate to stabilize the ferroelastic LS phase.  

We also find  a dimer phase, which results from the competition between antiferromagnetic and antiferroelastic interactions. This is equivalent to the dimer phase previously predicted for chains of d$^6$ complexes in the absence of SOC by Timm and Schollw\"ock \cite{timm}.

Natural questions arising from our work include: What effect do finite temperatures have of the phase diagram, especially do any additional phases occur at finite temperature? What happens in higher dimensional systems? What are the equivalent models for   coordination spheres with different symmetries or different fillings of the metallic ions? And how does the emergent physics of such models differ from that described here?

\begin{acknowledgments}
This work was supported by the Australian Research Council (DP230100139), the Australian Government Research
Training Program Scholarship, and MEXT Quantum Leap Flagship Program (MEXT Q-LEAP) Grant Number JPMXS0118069605.
\end{acknowledgments}

%\appendix

%\section{Additional figures}
%\label{sec:appendix}

% \begin{figure}
% \begin{centering}
% \includegraphics[width=\linewidth]{images/paper_figures/AFMappendix_no_shading.pdf}
% \caption{\label{fig:AFMappendix}
% Antiferroelastic:
% %Different phase transitions present for antiferroelastic coupling. 
% (a) Weak SOC $\lambda / J_I = 0.01$ and 
% (b) strong SOC $\lambda / J_I = 1.1$ for $\Delta G = 0$.
% (c) Weak SOC $\lambda / J_I = 0.5$ and spin-exchange $J_H / J_I = 0.2$, and 
% (d) strong SOC $\lambda / J_I = 1.5$ and spin-exchange $J_H / J_I = 2.6$.
% }
% \end{centering}
% \end{figure}

%

\bibliography{bibliography.bbl}

\end{document}